\documentclass[aps, prl,twocolumn, 10pt, showpacs, longbibliography, superscriptaddress]{revtex4-1}
\usepackage{graphicx}% Include figure files
\usepackage{lipsum}
\usepackage{dcolumn}% Align table columns on decimal point
\usepackage{bm}% bold math
\usepackage{color}
\usepackage{amsmath, amsthm, amssymb}
\usepackage[modulo]{lineno}%[switch]{lineno} 
\usepackage[usenames,dvipsnames]{xcolor}
\usepackage[normalem]{ulem}
\newcommand{\ba}{\begin{eqnarray}}
\newcommand{\ea}{\end{eqnarray}}

\usepackage[colorlinks=true, urlcolor=violet, linkcolor=blue, citecolor=red, hyperindex=true, linktocpage=true]{hyperref}

\usepackage{float}
\floatstyle{boxed}
\graphicspath{ {Figures/} }

\begin{document}

\title{Miniaturizing transmon qubits using van der Waals materials}

\author{Abhinandan Antony}
  \affiliation{Department of Mechanical Engineering, Columbia University, New York, NY 10027, USA}
\author{Martin V. Gustafsson}
 \affiliation{Raytheon BBN Technologies, Quantum Engineering and Computing Group, Cambridge, Massachusetts 02138, USA}
 \author{Guilhem J. Ribeill}
  \affiliation{Raytheon BBN Technologies, Quantum Engineering and Computing Group, Cambridge, Massachusetts 02138, USA}
\author{Matthew Ware}
  \affiliation{Raytheon BBN Technologies, Quantum Engineering and Computing Group, Cambridge, Massachusetts 02138, USA}
 \author{Anjaly Rajendran}
 \affiliation{Department of Electrical Engineering, Columbia University, New York, NY 10027, USA}
 \author{Luke C. G. Govia}
  \affiliation{Raytheon BBN Technologies, Quantum Engineering and Computing Group, Cambridge, Massachusetts 02138, USA}
\author{Thomas A. Ohki}
\affiliation{Raytheon BBN Technologies, Quantum Engineering and Computing Group, Cambridge, Massachusetts 02138, USA}
\author{Takashi Taniguchi}
\affiliation{International Center for Materials Nanoarchitectonics, National Institute for Materials Science,  1-1 Namiki, Tsukuba 305-0044, Japan}
\author{Kenji Watanabe}
\affiliation{Research Center for Functional Materials, National Institute for Materials Science, 1-1 Namiki, Tsukuba 305-0044, Japan}
\author{James Hone}
 \affiliation{Department of Mechanical Engineering, Columbia University, New York, NY 10027, USA}
\author{Kin Chung Fong}
\email{kc.fong@rtx.com}
\affiliation{Raytheon BBN Technologies, Quantum Engineering and Computing Group, Cambridge, Massachusetts 02138, USA}
%% Notice placement of commas and superscripts and use of &
%% in the author \listoffigures

\date{\today}
\begin{abstract}
Quantum computers can potentially achieve an exponential speedup versus classical computers on certain computational tasks, recently demonstrated in superconducting qubit processors. However, the capacitor electrodes that comprise these qubits must be large in order to avoid lossy dielectrics. This tactic hinders scaling by increasing parasitic coupling among circuit components, degrading individual qubit addressability, and limiting the spatial density of qubits. Here, we take advantage of the unique properties of van der Waals (vdW) materials to reduce the qubit area by $>1000$ times while preserving the capacitance without increasing substantial loss. Our qubits combine conventional aluminum-based Josephson junctions with parallel-plate capacitors composed of crystalline layers of superconducting niobium diselenide and insulating hexagonal-boron nitride. We measure a vdW transmon $T_1$ relaxation time of 1.06 $\mu$s, demonstrating a path to achieve high-qubit-density quantum processors with long coherence times, and the broad utility of layered heterostructures in low-loss, high-coherence quantum devices.
\end{abstract}

% \begin{tocentry}
% \includegraphics{nanoletters_toc.pdf}
% \end{tocentry}

%\pacs{65.80.Ck, 68.65.-k, and 07.57.Kp}

\maketitle

%\linenumbers
%\setstretch{2}
%\doublespacing

\begin{figure*}[t]
\includegraphics[width=1.5\columnwidth]{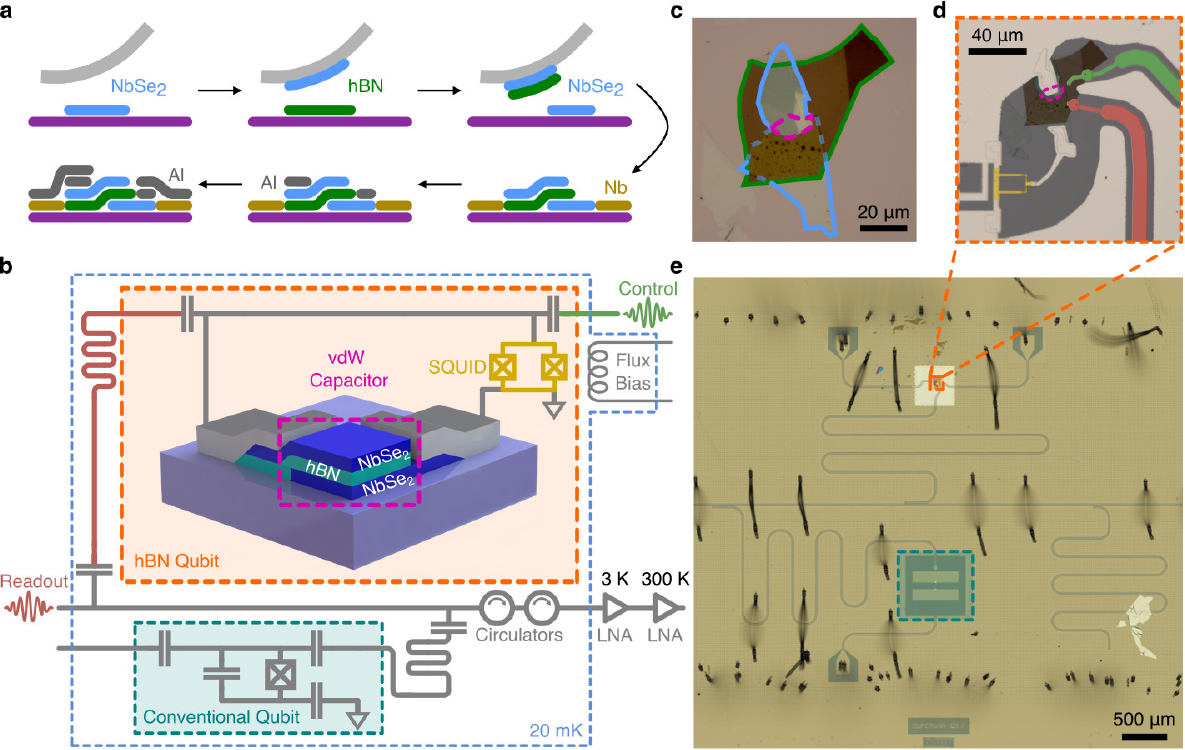}
\caption{Fabrication of vdW transmon and measurement schematics.  (a) Fabrication procedures for the vdW parallel-plate capacitor (PPC). In clockwise direction: dry polymer-based stacking of vdW heterostructure by sequential pick-up of NbSe$_2$, hBN, and NbSe$_2$ flakes. Transfer of the heterostructure to a void within the pre-patterned Nb layer on a silicon substrate,   deposition of Al contacts, and deposition of JJs and interconnects. (b) Schematic circuit diagram showing our frequency-tunable qubit (orange box) and fixed-frequency conventional qubit (teal box) as a control. Both qubits are coupled to coplanar-waveguide resonators for readout through a common bus line with low noise amplifiers (LNA) after circulators. (c) Optical micrograph of the NbSe$_2$-hBN-NbSe$_2$ heterostructure used to make the PPC (pink dashed line) with an area of 109 $\mu$m$^2$ and hBN thickness of 35 nm. (d) Optical micrograph of the vdW transmon formed by the SQUID loop (yellow) shunted by the vdW capacitor with capacitive coupling to the readout resonator (red) and control line (green). (e) Optical micrograph of both qubits in the same field of view, illustrating the miniaturization of transmon qubit by the vdW capacitor.}
\label{fig:1}\end{figure*}

Superconducting qubits form an attractive platform to build scalable quantum computers, as recently demonstrated in quantum processors consisting of large arrays of qubits \cite{Arute:2019fg}. Achieving fault-tolerant quantum computation would provide a transformative technology in quantum simulation, chemistry, and optimization \cite{Cao:2019jn, Berry:2020iy}. However, the industry-standard transmon qubit consists of Josephson junctions (JJ) shunted by capacitors with a large footprint of $\sim 10^5\mu$m$^2$. In order to acquire a sufficient capacitance, these capacitors are made of large planar electrodes without dielectric materials that can host two-level systems (TLSs), which are a major source of decoherence \cite{Muller:2019ie}. While brute-force scaling using conventional transmons remains possible \cite{Gambetta:bn}, there is growing interest in alternative approaches that could lay the foundation of  next-generation quantum processors. Approaches such as protected qubits can decouple from environmental noise \cite{Gyenis:2021em}, but improving the constituent materials can dramatically impact qubit technology by reducing the footprint and enhancing qubit coherence. As such, new material platforms for qubits are now actively explored \cite{Casparis:2018il, Lee:2019gn, Wisbey:2019je, McRae:2021bo, Place:2021hp, Mamin:2021ee, Antony:2021ws}.

Van der Waals (vdW) heterostructures are a promising material platform for quantum devices. Cleaved from pure bulk crystals, the surfaces of these heterostructures are pristine and atomically flat. They can be assembled into stacks with their layers held together by weak bonds without straining the crystals \cite{Wang:2013ch}. This architecture allows for diverse properties of various vdW materials, such as gate-tunability \cite{Schmidt:2018ji, Wang:2019dy}, high kinetic inductance \cite{Seifert:2020gy}, and giant thermal response to microwave photons \cite{Lee:2020ci, Kokkoniemi:2020dk}, which can be exploited for potential quantum applications \cite{Liu:2019if}. Here, we take advantage of the high-quality vdW heterostructures to fabricate compact vdW parallel-plate capacitors (PPC) using superconducting niobium diselenide (NbSe$_2$) and insulating hexagonal boron nitride (hBN). We observe quantum coherence of a transmon qubit that uses a vdW PPC as the shunt capacitor. Our result goes against the current trend of requiring large-footprint shunt capacitors in transmons \cite{Gambetta:bn}, and demonstrates the potential of using crystalline materials to improve qubit coherence.

Fig. 1a shows schematically the fabrication process of the vdW PPC \cite{Antony:2021ws} (details in Supplementary Information). It is different from the conventional process for making high-quality vdW heterostructures \cite{Wang:2013ch} because qubits operate in the single-photon regime, where any dissipation or loss channel introduced by the materials or fabrication process can destroy quantum information. Therefore, we assemble each heterostructure within an inert atmosphere to avoid oxidation and contamination of the interfaces. Then, the vdW heterostructure is placed onto the qubit substrate with pre-patterned niobium resonators. To make a vdW transmon, we connect the vdW PPC in parallel with two Al-AlO$_x$-Al JJs that form a superconducting quantum interference device (SQUID) loop for tuning of the qubit frequency with a magnetic flux generated by a bias current through an off-chip coil. A conventional qubit formed using a planar Nb capacitor and a single JJ is created at the same time for comparison. Fig. 1b schematically shows these elements, with the transmons capacitively coupled to the readout resonator (red) and drive line (green). Fig. 1c and 1d show optical micrographs of the assembled vdW PPC (109 $\mu$m$^2$ in area and 35 nm in thickness) and the vdW transmon, respectively. The optical image in Fig. 1e shows the dramatic size reduction enabled by the vdW PPC, which is about 1400 times smaller than the conventional planar shunt capacitor.  For this study, we fabricated and measured two vdW transmons. We fully characterize one of them and present the data in the main text.

We perform spectroscopy experiments to characterize the coupling of the vdW qubit to its readout cavity. With high probing power, the cavity is unperturbed by the qubit and resonates at 6.9073 GHz with a linewidth of 290 kHz (Fig. 2a). Measuring under low power, we find the strong coupling with the qubit dispersively shifts the cavity frequency by 1.00 MHz \cite{Blais:2021be}. Fig. 2b shows how the flux-bias current modulates the cavity frequency. We perform qubit spectroscopy by applying a control signal at various frequencies to the qubit while probing the cavity nearly on resonance. Fig. 2c shows the transition frequency, $\omega_{01}/2\pi$, between the ground, $|0\rangle$, and excited, $|1\rangle$, states of the qubit at 5.2815 GHz. Using the two-photon transition frequency between the $|0\rangle$ and $|2\rangle$ states, i.e. $\omega_{02}/2\pi =$ 5.216 GHz, we find the qubit anharmonicity, $\alpha/2\pi$, = -131 MHz. $\alpha$ agrees with the charging energy of the qubit, $E_C$ ($-\alpha = E_C/h$ for a transmon with $h$ being the Planck constant), calculated using the PPC formula with an hBN dielectric constant of 4.4. Fig. 2d shows $\omega_{01}/2\pi$ versus flux-bias current. It has a maximum at $\approx$0.2 mA, which is consistent with the flux dependence of the readout resonator in Fig. 2a and represents the sweet spot where the transmon is the least susceptible to flux noise. In 5-$\mu$A steps of flux-bias current, we observe no anticrossing in the qubit spectroscopy due to the coupling of the vdW transmon with coherent TLSs. From the dispersive shift and detuning from the readout cavity, $\Delta$, of 1.6258 GHz at the sweet spot, we find the qubit-resonator coupling, $g/2\pi$, = 40.3 MHz. The Josephson energy, $E_J/h$, calculated from $\omega_{01}$, is 28.0 GHz. Our qubit is in the transmon regime with $E_J/E_C \simeq 214$.

\begin{figure}
\includegraphics[width=\columnwidth]{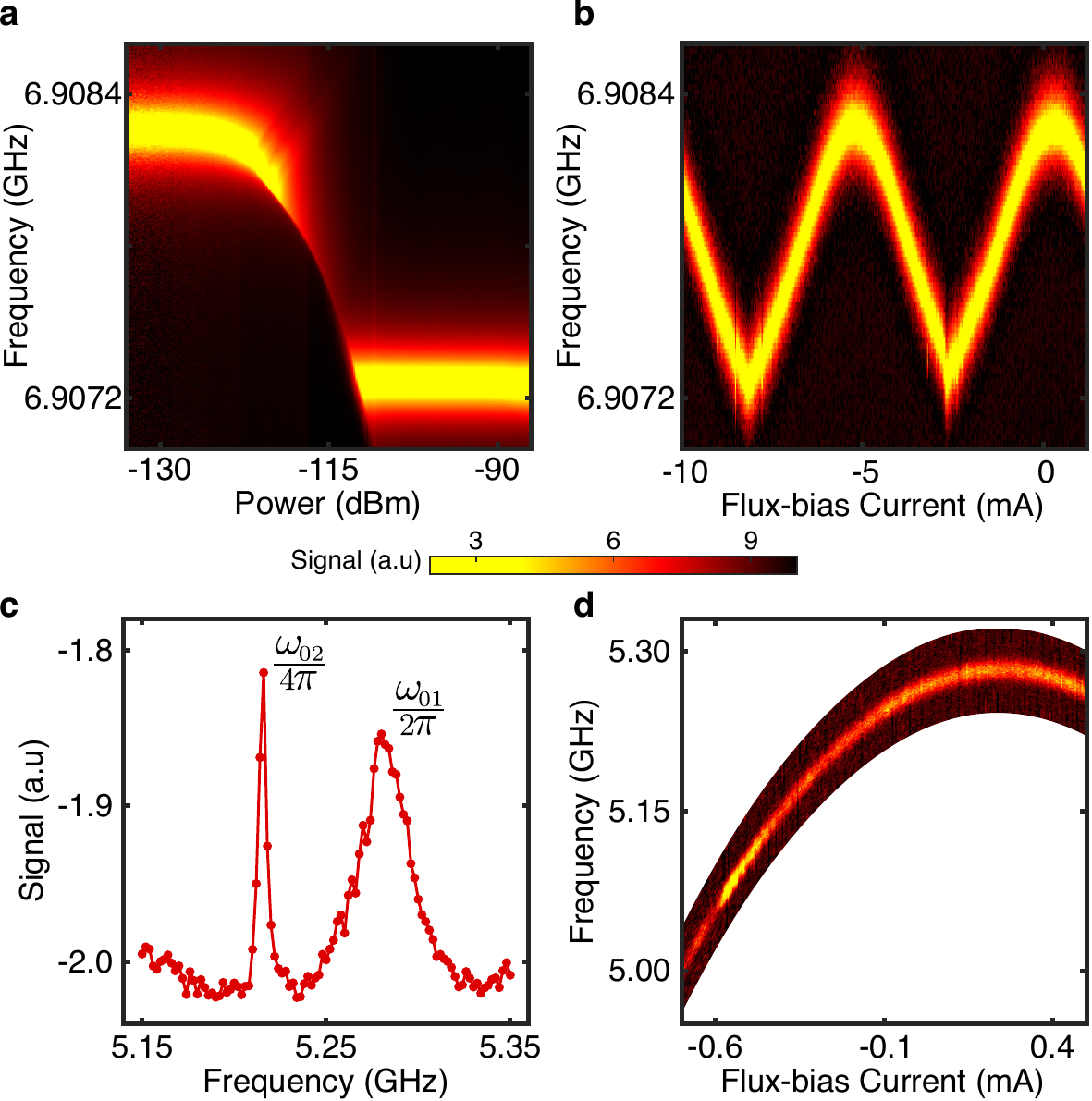}
\caption{Spectroscopy of a vdW transmon (a) Readout-cavity resonance as a function of applied microwave power at the sweet spot of flux bias. (b) Readout-cavity resonance versus flux-bias current. (c) High power spectroscopy shows the $\omega_{01}$ and two-photon $\omega_{02}/2$ qubit transitions at the sweet spot of flux bias. (d) Qubit frequency, $\omega_{01}/2\pi$, as a function of flux tuning.}
\label{fig:spec}\end{figure}

\begin{figure*}[t]
\includegraphics[width=1.5\columnwidth]{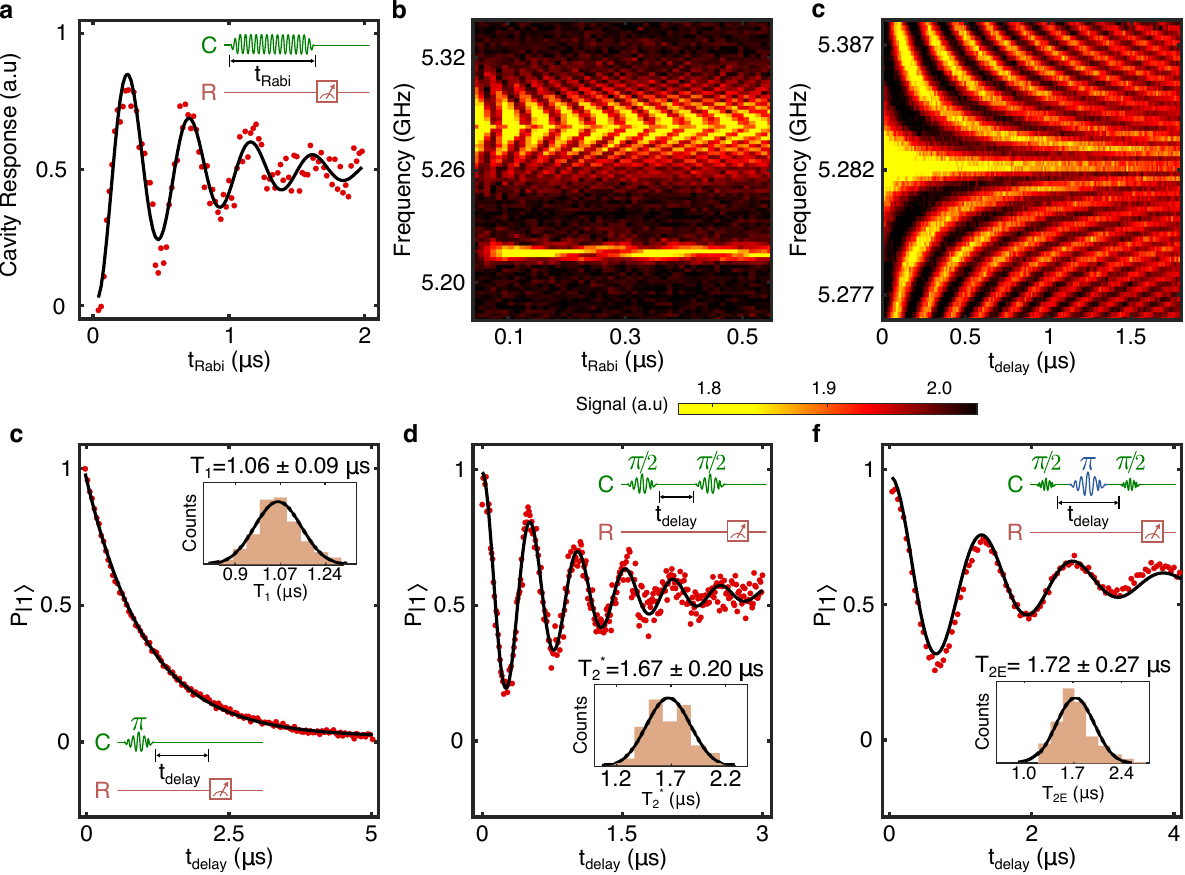}
\caption{Quantum coherence of the vdW transmon at the sweet spot of flux bias. (a) Rabi oscillation vs. the pulse width, $t_{Rabi}$ near the qubit frequency. (b) Rabi oscillations at a high power for different excitation frequencies. (c) $T_1$ energy relaxation time measurement with the pulse control and measurement sequence (inset). The exponential fit yields the $T_1$ relaxation time. The inset histogram shows data from 8 hours of repeated measurements. d) Ramsey experiment shows $T_2^*$ dephasing time. (e) Ramsey fringes at different drive tone frequencies. (f)  Hahn echo experiment to obtain the echo dephasing time, $T_{2E}$.}
\label{fig:coherence}\end{figure*}

We demonstrate the quantum coherence of the  vdW transmon by performing Rabi experiments (see Fig. 3a). By driving the qubit with a resonant pulse for a time, $t_{Rabi}$, the vdW transmon Rabi oscillates coherently on the Bloch sphere between the ground, $|0\rangle$, and excited, $|1\rangle$, states before being measured using the dispersive shift of the resonator \cite{Blais:2021be}. The oscillation frequency, $\omega_{\rm Rabi}/2\pi$ is set by the drive strength, $\Omega$, and drive detuning from the qubit frequency, $\delta\omega_d$, as  $\omega_{\rm Rabi} = \sqrt{\Omega^2+\delta\omega_d^2}$. When performed at different excitation frequencies and a high pulse power, we obtain the data as shown in Fig. 3b. We quantify the qubit lifetime by measuring the $T_1$ energy relaxation time at the sweet spot (Fig. 3c). After applying an 80-ns-long $\pi$-pulse to excite the qubit from $|0\rangle$ to $|1\rangle$ and waiting for a time delay, $t_{delay}$, the qubit state is read out. We repeatedly measure the exponential decay of excited state probability, $P_{|1\rangle}$, to quantify the mean and fluctuations of $T_1$ over a period of 8 hours. The histogram (see inset) shows $T_1$ of 1.06 $\mu$s $\pm 0.09 \mu$s (standard deviation). Next, we quantify the qubit coherence by measuring the dephasing time $T_2^*$ during a Ramsey experiment. Fig. 3d shows the qubit readout signal after the Ramsey pulse sequence of $\pi/2$-$t_{delay}$-$\pi/2$. The oscillation is due to $\delta\omega_d$. We fit the data to a decay oscillation and obtain $T_2^*$ of $1.67\pm 0.20 \mu$s. Fig. 3e shows Ramsey oscillations at different qubit drive frequencies. Finally, we probe for evidence of excessive low frequency noise that may dephase the vdW transmon using a Hahn echo pulse sequence, which contains an additional $\pi$-pulse in the middle of the Ramsey sequence. Fig. 3f shows the qubit readout signal after the Hahn echo sequence. We obtain the echo dephasing time, $T_{2E}$, of $1.72 \pm 0.27 \mu$s, that is nearly energy-relaxation limited, i.e. $T_{2E} \approx 2 T_{1}$.

We can compare the performance of our vdW transmon with the conventional one fabricated on the same substrate. The $T_1$ and $T_2^*$ of the conventional transmon are $11.5\pm 0.4 \mu$s and $10.5\pm 0.6 \mu$s, respectively. Since both transmons are subjected to the same fabrication process and measurement setup environment, this result suggests that the JJs are not the limiting factor of the quantum coherence of our vdW transmon. On the other hand, based on finite-element simulations, only 87\% of the capacitive energy of our vdW transmon design is confined internally to the PPC. This could result in degraded qubit performance through a finite qubit coupling with residue, external to the PPC, from the vdW stacking process or Al lift-off. If we assume the energy relaxation is entirely caused by the capacitive loss and use the equation,  $1/T_1 = \omega_{01}\tan{\delta}$, we can estimate the upper bound of the loss tangent, $\tan{\delta}$, of the vdW PPC to be $2.83\times 10^{-5}$. This value is not the intrinsic $\tan{\delta}$ of the hBN, but is approaching the measured values from PPCs using silicon as the dielectric ($\approx 10^{-6}$) \cite{Patel:2013ko, Sandberg:2013dk}. To extend qubit $T_1$ relaxation, we will need to improve the fabrication process and further concentrate the electric field within the pristine interior of the vdW PPC.

Recently, experiments have taken a similar direction using the intrinsic JJ PPC to make the merged-element transmon (MET) \cite{Zhao:2020jw, Mamin:2021ee}. While both the MET and vdW transmon accomplish miniaturization, the METs demonstrated so far use amorphous insulators which are known to harbor TLSs. As the vdW heterostructure can also function as a high-quality JJ \cite{Island:2016gz, Dvir:2018fj, Lee:2019gn}, one can envision a future  vdW-MET consisting of a crystalline vdW material as the tunnel barrier. Moreover, the thickness of the vdW insulator is defined by discrete atomic layers. This highly-uniform, crystalline tunnel barrier can potentially allow for a reduced TLS defect density and precise control of qubit frequencies, which can address the frequency-crowding problem \cite{Brink:2018wu}. In the future, we plan to extend our result to further miniaturize quantum processors by reducing footprints of circuit components using, e.g. compact lumped-element resonators.

During the preparation of this manuscript, we became aware of a complementary work \cite{Wang:2021uq}.

\textbf{Acknowledgements} We thank L.~Ranzani for useful discussions. Most of this work was supported by Army Research Office under Contract Number W911NF-18-C-0044. Development of heterostructure assembly techniques at Columbia was supported by the NSF MRSEC program (DMR-2011736). A.A. thanks the supplemental support from QISE-NET under NSF DMR-1747426. K.W. and T.T. acknowledge support from the Elemental Strategy Initiative conducted by the MEXT, Japan (Grant Number JPMXP0112101001) and  JSPS KAKENHI (Grant Numbers JP19H05790 and JP20H00354).

\textbf{Supporting Information}
Supporting material containing details about the fabrication method, qubit parameters, measurements of a conventional qubit as well as a second vdW qubit, is available free of charge via the internet at https://pubs.acs.org/doi/10.1021/acs.nanolett.1c04160.

%\pagebreak
%\bibliographystyle{Science}
%\bibliographystyle{naturemag}
%\bibliography{QubitV1}

\begin{thebibliography}{10}

\bibliographystyle{achemso}
\expandafter\ifx\csname url\endcsname\relax
  \def\url#1{\texttt{#1}}\fi
\expandafter\ifx\csname urlprefix\endcsname\relax\def\urlprefix{URL }\fi
\providecommand{\bibinfo}[2]{#2}
\providecommand{\eprint}[2][]{\url{#2}}

\bibitem{Arute:2019fg}
\bibinfo{author}{Arute, F.} \emph{et~al.}
\newblock \bibinfo{title}{{Quantum supremacy using a programmable
  superconducting processor}}.
\newblock \emph{\bibinfo{journal}{Nature}}
  \textbf{\bibinfo{volume}{574}}, \bibinfo{pages}{505--510}
  (\bibinfo{year}{2019}).

\bibitem{Cao:2019jn}
\bibinfo{author}{Cao, Y.} \emph{et~al.}
\newblock \bibinfo{title}{{Quantum Chemistry in the Age of Quantum Computing}}.
\newblock \emph{\bibinfo{journal}{Chem. Rev.}}
  \textbf{\bibinfo{volume}{119}}, \bibinfo{pages}{10856} (\bibinfo{year}{2019}).

\bibitem{Berry:2020iy}
\bibinfo{author}{Berry, D.~W.}, \bibinfo{author}{Childs, A.~M.},
  \bibinfo{author}{Su, Y.}, \bibinfo{author}{Wang, X.} \&
  \bibinfo{author}{Wiebe, N.}
\newblock \bibinfo{title}{{Time-dependent Hamiltonian simulation with
  $L^1$-norm scaling}}.
\newblock \emph{\bibinfo{journal}{Quantum}} \textbf{\bibinfo{volume}{4}},
  \bibinfo{pages}{254} (\bibinfo{year}{2020}).

\bibitem{Muller:2019ie}
\bibinfo{author}{M{\"u}ller, C.}, \bibinfo{author}{Cole, J.~H.} \&
  \bibinfo{author}{Lisenfeld, J.}
\newblock \bibinfo{title}{{Towards understanding two-level-systems in amorphous
  solids: insights from quantum circuits}}.
\newblock \emph{\bibinfo{journal}{Rep. Prog. Phys.}}
  \textbf{\bibinfo{volume}{82}}, \bibinfo{pages}{124501}
  (\bibinfo{year}{2019}).

\bibitem{Gambetta:bn}
\bibinfo{author}{Gambetta, J.~M.} \emph{et~al.}
\newblock \bibinfo{title}{{Investigating Surface Loss Effects in
  Superconducting Transmon Qubits}}.
\newblock \emph{\bibinfo{journal}{IEEE Trans. Appl. Supercond.}}
  \textbf{\bibinfo{volume}{27}}, \bibinfo{pages}{1--5}
  (\bibinfo{year}{2017}).

\bibitem{Gyenis:2021em}
\bibinfo{author}{Gyenis, A.} \emph{et~al.}
\newblock \bibinfo{title}{{Experimental Realization of a Protected
  Superconducting Circuit Derived from the 0{\textendash}$\pi$ Qubit}}.
\newblock \emph{\bibinfo{journal}{PRX Quantum}}
  \textbf{\bibinfo{volume}{2}}, \bibinfo{pages}{010339}
  (\bibinfo{year}{2021}).

\bibitem{Casparis:2018il}
\bibinfo{author}{Casparis, L.} \emph{et~al.}
\newblock \bibinfo{title}{{Superconducting gatemon qubit based on a
  proximitized two-dimensional electron gas}}.
\newblock \emph{\bibinfo{journal}{Nat. Nanotechnol.}}
  \textbf{\bibinfo{volume}{13}}, \bibinfo{pages}{915--919}
  (\bibinfo{year}{2018}).

\bibitem{Lee:2019gn}
\bibinfo{author}{Lee, K.-H.} \emph{et~al.}
\newblock \bibinfo{title}{{Two-Dimensional Material Tunnel Barrier for
  Josephson Junctions and Superconducting Qubits}}.
\newblock \emph{\bibinfo{journal}{Nano Lett.}} \textbf{\bibinfo{volume}{19}},
  \bibinfo{pages}{8287} (\bibinfo{year}{2019}).

\bibitem{Wisbey:2019je}
\bibinfo{author}{Wisbey, D.~S.} \emph{et~al.}
\newblock \bibinfo{title}{{Dielectric Loss of Boron-Based Dielectrics on
  Niobium Resonators}}.
\newblock \emph{\bibinfo{journal}{J. Low Temp. Phys.}}
  \textbf{\bibinfo{volume}{195}}, \bibinfo{pages}{474--486}
  (\bibinfo{year}{2019}).

\bibitem{McRae:2021bo}
\bibinfo{author}{McRae, C. R.~H.} \emph{et~al.}
\newblock \bibinfo{title}{{Cryogenic microwave loss in epitaxial Al/GaAs/Al
  trilayers for superconducting circuits}}.
\newblock \emph{\bibinfo{journal}{J. Appl. Phys.}}
  \textbf{\bibinfo{volume}{129}}, \bibinfo{pages}{025109}
  (\bibinfo{year}{2021}).

\bibitem{Place:2021hp}
\bibinfo{author}{Place, A. P.~M.} \emph{et~al.}
\newblock \bibinfo{title}{{New material platform for superconducting transmon
  qubits with coherence times exceeding 0.3 milliseconds}}.
\newblock \emph{\bibinfo{journal}{Nat. Comm.}}
  \textbf{\bibinfo{volume}{12}}, \bibinfo{pages}{1--6}
  (\bibinfo{year}{2021}).

\bibitem{Mamin:2021ee}
\bibinfo{author}{Mamin, H.~J.} \emph{et~al.}
\newblock \bibinfo{title}{{Merged-Element Transmons: Design and Qubit
  Performance}}.
\newblock \emph{\bibinfo{journal}{Phys. Rev. Appl.}}
  \textbf{\bibinfo{volume}{16}}, \bibinfo{pages}{024023}
  (\bibinfo{year}{2021}).

\bibitem{Antony:2021ws}
\bibinfo{author}{Antony, A.} \emph{et~al.}
\newblock \bibinfo{title}{{Making high-quality quantum microwave devices with
  van der Waals superconductors.}}
\newblock (\bibinfo{year}{2021})  \bibinfo{pages}{2107.09147}
  \emph{\bibinfo{journal}{arXiv}}
\newblock \bibinfo{url}{https://arxiv.org/abs/2109.02824 (Accessed Nov 10, 2021)}.

\bibitem{Wang:2013ch}
\bibinfo{author}{Wang, L.} \emph{et~al.}
\newblock \bibinfo{title}{{One-Dimensional Electrical Contact to a
  Two-Dimensional Material}}.
\newblock \emph{\bibinfo{journal}{Science}}
  \textbf{\bibinfo{volume}{342}}, \bibinfo{pages}{614--617}
  (\bibinfo{year}{2013}).

\bibitem{Schmidt:2018ji}
\bibinfo{author}{Schmidt, F.~E.}, \bibinfo{author}{Jenkins, M.~D.},
  \bibinfo{author}{Watanabe, K.}, \bibinfo{author}{Taniguchi, T.} \&
  \bibinfo{author}{Steele, G.~A.}
\newblock \bibinfo{title}{{A ballistic graphene superconducting microwave
  circuit}}.
\newblock \emph{\bibinfo{journal}{Nat. Comm.}}
  \textbf{\bibinfo{volume}{9}}, \bibinfo{pages}{4069}
  (\bibinfo{year}{2018}).

\bibitem{Wang:2019dy}
\bibinfo{author}{Wang, J. I.-J.} \emph{et~al.}
\newblock \bibinfo{title}{{Coherent control of a hybrid superconducting circuit
  made with graphene-based van der Waals heterostructures}}.
\newblock \emph{\bibinfo{journal}{Nat. Nanotechnol.}}
  \textbf{\bibinfo{volume}{14}}, \bibinfo{pages}{120--125}
  (\bibinfo{year}{2019}).

\bibitem{Seifert:2020gy}
\bibinfo{author}{Seifert, P.} \emph{et~al.}
\newblock \bibinfo{title}{{Magic-Angle Bilayer Graphene Nanocalorimeters:
  Toward Broadband, Energy-Resolving Single Photon Detection}}.
\newblock \emph{\bibinfo{journal}{Nano Lett.}} \textbf{\bibinfo{volume}{20}},
  \bibinfo{pages}{3459} (\bibinfo{year}{2020}).

\bibitem{Lee:2020ci}
\bibinfo{author}{Lee, G.-H.} \emph{et~al.}
\newblock \bibinfo{title}{{Graphene-based Josephson junction microwave
  bolometer}}.
\newblock \emph{\bibinfo{journal}{Nature}}
  \textbf{\bibinfo{volume}{586}}, \bibinfo{pages}{42--46}
  (\bibinfo{year}{2020}).

\bibitem{Kokkoniemi:2020dk}
\bibinfo{author}{Kokkoniemi, R.} \emph{et~al.}
\newblock \bibinfo{title}{{Bolometer operating at the threshold for circuit
  quantum electrodynamics}}.
\newblock \emph{\bibinfo{journal}{Nature}}
  \textbf{\bibinfo{volume}{586}}, \bibinfo{pages}{47--51}
  (\bibinfo{year}{2020}).

\bibitem{Liu:2019if}
\bibinfo{author}{Liu, X.} \& \bibinfo{author}{Hersam, M.~C.}
\newblock \bibinfo{title}{{2D materials for quantum information science}}.
\newblock \emph{\bibinfo{journal}{Nat. Rev. Mater.}}
  \textbf{\bibinfo{volume}{4}}, \bibinfo{pages}{669--684}
  (\bibinfo{year}{2019}).

\bibitem{Blais:2021be}
\bibinfo{author}{Blais, A.}, \bibinfo{author}{Grimsmo, A.~L.},
  \bibinfo{author}{Girvin, S.~M.} \& \bibinfo{author}{Wallraff, A.}
\newblock \bibinfo{title}{{Circuit quantum electrodynamics}}.
\newblock \emph{\bibinfo{journal}{Rev. Mod. Phys.}}
  \textbf{\bibinfo{volume}{93}}, \bibinfo{pages}{025005}
  (\bibinfo{year}{2021}).

\bibitem{Patel:2013ko}
\bibinfo{author}{Patel, U.} \emph{et~al.}
\newblock \bibinfo{title}{{Coherent Josephson phase qubit with a single crystal
  silicon capacitor}}.
\newblock \emph{\bibinfo{journal}{Appl. Phys. Lett.}}
  \textbf{\bibinfo{volume}{102}}, \bibinfo{pages}{012602}
  (\bibinfo{year}{2013}).

\bibitem{Sandberg:2013dk}
\bibinfo{author}{Sandberg, M.} \emph{et~al.}
\newblock \bibinfo{title}{{Radiation-suppressed superconducting quantum bit in
  a planar geometry}}.
\newblock \emph{\bibinfo{journal}{Appl. Phys. Lett.}}
  \textbf{\bibinfo{volume}{102}}, \bibinfo{pages}{072601}
  (\bibinfo{year}{2013}).

\bibitem{Zhao:2020jw}
\bibinfo{author}{Zhao, R.} \emph{et~al.}
\newblock \bibinfo{title}{{Merged-Element Transmon}}.
\newblock \emph{\bibinfo{journal}{Phys. Rev. Appl.}}
  \textbf{\bibinfo{volume}{14}}, \bibinfo{pages}{064006}
  (\bibinfo{year}{2020}).

\bibitem{Island:2016gz}
\bibinfo{author}{Island, J.~O.}, \bibinfo{author}{Steele, G.~A.},
  \bibinfo{author}{van~der Zant, H. S.~J.} \&
  \bibinfo{author}{Castellanos-Gomez, A.}
\newblock \bibinfo{title}{{Thickness dependent interlayer transport in vertical
  MoS2 Josephson junctions}}.
\newblock \emph{\bibinfo{journal}{2D Materials}}
  \textbf{\bibinfo{volume}{3}}, \bibinfo{pages}{031002}
  (\bibinfo{year}{2016}).

\bibitem{Dvir:2018fj}
\bibinfo{author}{Dvir, T.} \emph{et~al.}
\newblock \bibinfo{title}{{Spectroscopy of bulk and few-layer superconducting
  NbSe 2 with van der Waals tunnel junctions}}.
\newblock \emph{\bibinfo{journal}{Nat. Comm.}}
  \textbf{\bibinfo{volume}{9}}, \bibinfo{pages}{1--6}
  (\bibinfo{year}{2018}).

\bibitem{Brink:2018wu}
\bibinfo{author}{Brink, M.}, \bibinfo{author}{Chow, J.~M.},
  \bibinfo{author}{Hertzberg, J.}, \bibinfo{author}{Magesan, E.} \&
  \bibinfo{author}{Rosenblatt, S.}
\newblock \bibinfo{title}{{Device challenges for near term superconducting
  quantum processors: frequency collisionsDevice challenges for near term
  superconducting quantum processors: frequency collisions}}.
\newblock \emph{\bibinfo{journal}{IEEE International Electron Devices Meeting
  IEDM}} \bibinfo{pages}{6.1.1} (\bibinfo{year}{2018}).

\bibitem{Wang:2021uq}
\bibinfo{author}{Wang, J. I.~J.} \emph{et~al.}
\newblock \bibinfo{title}{{Hexagonal Boron Nitride (hBN) as a Low-loss
  Dielectric for Superconducting Quantum Circuits and Qubits.}}
\newblock (\bibinfo{year}{2021})  \bibinfo{pages}{2107.09147}
  \emph{\bibinfo{journal}{arXiv}}
\newblock \bibinfo{url}{https://arxiv.org/abs/2109.00015 (Accessed Nov 10, 2021)}.
\end{thebibliography}

\end{document}